\documentclass[12pt]{article}
\usepackage{amssymb}
\usepackage{graphicx}
\usepackage{xcolor}

\definecolor{primary}{RGB}{0, 139, 139}      
\definecolor{secondary}{RGB}{180, 40, 40}   
\definecolor{accent}{RGB}{40, 120, 120}      
\usepackage[colorlinks=true]{hyperref}
\hypersetup{linkcolor=primary,citecolor=accent,urlcolor=secondary}

\begin{document}
\newcommand{\beq}{\begin{equation}}
\newcommand{\eeq}{\end{equation}}
\newcommand{\beqa}{\begin{eqnarray}}
\newcommand{\eeqa}{\end{eqnarray}}
\newcommand{\beqar}{\begin{eqnarray*}}
\newcommand{\eeqar}{\end{eqnarray*}}
\newcommand{\al}{\alpha}
\newcommand{\be}{\beta}
\newcommand{\del}{\delta}
\newcommand{\D}{\Delta}
\newcommand{\eps}{\epsilon}
\newcommand{\ga}{\gamma}
\newcommand{\Ga}{\Gamma}
\newcommand{\ka}{\kappa}
\newcommand{\nn}{\nonumber}
\newcommand{\inn}{\!\cdot\!}
\newcommand{\h}{\eta}
\newcommand{\ii}{\iota}
\newcommand{\kk}{\varphi}
\newcommand\F{{}_3F_2}
\newcommand{\la}{\lambda}
\newcommand{\La}{\Lambda}
\newcommand{\na}{\prt}
\newcommand{\Om}{\Omega}
\newcommand{\om}{\omega}
\newcommand{\p}{\Phi}
\newcommand{\sig}{\sigma}
\renewcommand{\t}{\theta}
\newcommand{\z}{\zeta}
\newcommand{\ssc}{\scriptscriptstyle}
\newcommand{\eg}{{\it e.g.,}\ }
\newcommand{\ie}{{\it i.e.,}\ }
\newcommand{\labell}[1]{\label{#1}} 
\newcommand{\reef}[1]{(\ref{#1})}
\newcommand\prt{\partial}
\newcommand\veps{\varepsilon}
\newcommand{\pol}{\varepsilon}
\newcommand\vp{\varphi}
\newcommand\ls{\ell_s}
\newcommand\cF{{\cal F}}
\newcommand\cA{{\cal A}}
\newcommand\cS{{\cal S}}
\newcommand\cT{{\cal T}}
\newcommand\cV{{\cal V}}
\newcommand\cL{{\cal L}}
\newcommand\cM{{\cal M}}
\newcommand\cN{{\cal N}}
\newcommand\cG{{\cal G}}
\newcommand\cK{{\cal K}}
\newcommand\cH{{\cal H}}
\newcommand\cI{{\cal I}}
\newcommand\cJ{{\cal J}}
\newcommand\cl{{\iota}}
\newcommand\cP{{\cal P}}
\newcommand\cQ{{\cal Q}}
\newcommand\cg{{\tilde {{\cal G}}}}
\newcommand\cR{{\cal R}}
\newcommand\cB{{\cal B}}
\newcommand\cO{{\cal O}}
\newcommand\tcO{{\tilde {{\cal O}}}}
\newcommand\bz{\bar{z}}
\newcommand\bb{\bar{b}}
\newcommand\ba{\bar{a}}
\newcommand\bg{\bar{g}}
\newcommand\bc{\bar{c}}
\newcommand\bw{\bar{w}}
\newcommand\bX{\bar{X}}
\newcommand\bK{\bar{K}}
\newcommand\bA{\bar{A}}
\newcommand\bH{\bar{H}}
\newcommand\bF{\bar{F}}
\newcommand\bxi{\bar{\xi}}
\newcommand\bphi{\bar{\phi}}
\newcommand\bpsi{\bar{\psi}}
\newcommand\bprt{\bar{\prt}}
\newcommand\bet{\bar{\eta}}
\newcommand\btau{\bar{\tau}}
\newcommand\hF{\hat{F}}
\newcommand\hA{\hat{A}}
\newcommand\hT{\hat{T}}
\newcommand\htau{\hat{\tau}}
\newcommand\hD{\hat{D}}
\newcommand\hf{\hat{f}}
\newcommand\hK{\hat{K}}
\newcommand\hg{\hat{g}}
\newcommand\hp{\hat{\Phi}}
\newcommand\hi{\hat{i}}
\newcommand\ha{\hat{a}}
\newcommand\hb{\hat{b}}
\newcommand\hQ{\hat{Q}}
\newcommand\hP{\hat{\Phi}}
\newcommand\hS{\hat{S}}
\newcommand\hX{\hat{X}}
\newcommand\tL{\tilde{\cal L}}
\newcommand\hL{\hat{\cal L}}
\newcommand\tG{{\tilde G}}
\newcommand\tg{{\tilde g}}
\newcommand\tphi{{\widetilde \Phi}}
\newcommand\tPhi{{\widetilde \Phi}}
\newcommand\te{{\tilde e}}
\newcommand\tk{{\tilde k}}
\newcommand\tf{{\tilde f}}
\newcommand\tH{{\tilde H}}
\newcommand\ta{{\tilde a}}
\newcommand\tb{{\tilde b}}
\newcommand\tc{{\tilde c}}
\newcommand\td{{\tilde d}}
\newcommand\tm{{\tilde m}}
\newcommand\tmu{{\tilde \mu}}
\newcommand\tnu{{\tilde \nu}}
\newcommand\talpha{{\tilde \alpha}}
\newcommand\tbeta{{\tilde \beta}}
\newcommand\trho{{\tilde \rho}}
 \newcommand\tR{{\tilde R}}
\newcommand\teta{{\tilde \eta}}
\newcommand\tF{{\widetilde F}}
\newcommand\tK{{\tilde K}}
\newcommand\tE{{\widetilde E}}
\newcommand\tpsi{{\tilde \psi}}
\newcommand\tX{{\widetilde X}}
\newcommand\tD{{\widetilde D}}
\newcommand\tO{{\widetilde O}}
\newcommand\tS{{\tilde S}}
\newcommand\tB{{\tilde B}}
\newcommand\tA{{\widetilde A}}
\newcommand\tT{{\widetilde T}}
\newcommand\tC{{\widetilde C}}
\newcommand\tV{{\widetilde V}}
\newcommand\thF{{\widetilde {\hat {F}}}}
\newcommand\Tr{{\rm Tr}}
\newcommand\tr{{\rm tr}}
\newcommand\STr{{\rm STr}}
\newcommand\hR{\hat{R}}
\newcommand\M[2]{M^{#1}{}_{#2}}
\newcommand\MZ{\mathbb{Z}}
\newcommand\MR{\mathbb{R}}
\newcommand\bS{\textbf{ S}}
\newcommand\bI{\textbf{ I}}
\newcommand\bJ{\textbf{ J}}

\begin{titlepage}
\begin{center}

\vskip 0.5 cm
{\LARGE \bf 
Tachyon–massless couplings at  order $\alpha'$ \\  \vskip 0.25 cm in bosonic string theory 
} \\
\vskip 1.25 cm
 Mohammad R. Garousi \footnote{garousi@um.ac.ir}

\vskip 1 cm
{{\it Department of Physics, Faculty of Science, Ferdowsi University of Mashhad\\}{\it P.O. Box 1436, Mashhad, Iran}\\}
\vskip .1 cm
 \end{center}

\begin{abstract}

Recently, it has been proposed that the classical effective action of bosonic string theory should contain only couplings with an even number of tachyon fields. We employ the T-duality procedure to derive such couplings at order \(\alpha'\) (four-derivative order) for the tachyon and massless fields. We first observe that, through higher-derivative field redefinitions of the tachyon potential term, one can choose a scheme in which—apart from the tachyon potential itself—all remaining couplings involve only covariant derivatives of the tachyon and the massless fields. We then construct a minimal basis of such couplings at four-derivative order, consisting of 14 independent terms. Imposing T-duality reduces this set to 6 nonzero couplings, expressed in terms of two unfixed parameters. One of these parameters is determined by matching to the known effective action in the zero-tachyon limit, while the other is fixed by comparison with the sphere-level S-matrix element of four tachyon vertex operators. This latter comparison also establishes that the coefficient of the \(T^4\) term in the tachyon potential is positive. 
Remarkably, the tachyon mass term and this quartic term are consistent with a potential of the form \((2/\alpha')(-1+\cos T)\), which implies that the closed string tachyon condenses to the minimum of the potential at \(T = \pi\), thereby generating an Anti–de Sitter spacetime with cosmological constant \(\Lambda = -4/\alpha'\).

\end{abstract}

\end{titlepage}

\tableofcontents

\section{Introduction}

Free bosonic string theory is a quantum theory of fluctuations of a relativistic string propagating in Minkowski spacetime. Its spectrum contains a tachyon, massless states, and an infinite tower of massive states. The interactions of these states in spacetime are described perturbatively by the S-matrix, which is computed from world-sheet correlation functions of the corresponding vertex operators at sphere, torus, and higher-genus levels \cite{Becker:2007zj}. In the real world, however, the massive states are too heavy to be produced, and one therefore seeks a low-energy effective action in spacetime that describes the interactions among the massless fields  \cite{Scherk:1971xy,Scherk:1974ca}. The presence of the tachyon in the spectrum signals that bosonic string theory in Minkowski spacetime is unstable. To circumvent this issue, one typically extends the bosonic theory to superstring theory, which is stable in Minkowski spacetime \cite{Becker:2007zj,Gross:1986iv,Gross:1986mw}.

In this paper, we explore an alternative way to address the instability of the bosonic theory. Rather than focusing solely on the tachyon mass—which corresponds only to the quadratic term in the tachyon potential—we aim to determine the higher-order terms in the potential and investigate whether a stable minimum exists. If such a minimum exists, then the condensation of the closed string tachyon to that minimum could result in a stable spacetime, which would no longer be Minkowski. The condensation of the open string tachyon has been extensively discussed in \cite{Sen:1999mg,Harvey:2000na,Minahan:2000ff,Minahan:2000tf,Dasgupta:2000kk}. To determine the closed string tachyon potential in perturbative string theory, one may use the sigma-model approach, which is based on the conformal symmetry of the world-sheet theory \cite{Tseytlin:2000mt,Tseytlin:1991bu,Banks:1991sg}. Alternatively, one may employ the S-matrix method together with the spacetime symmetries of the bosonic string. In addition to general covariance, the effective action of the bosonic theory is required to be invariant under T-duality transformations after reduction on a circle \cite{Becker:2007zj,Garousi:2017fbe}, which is a reflection of the conformal symmetry of the world-sheet theory in spacetime \cite{Buscher:1987sk,Rocek:1991ps}.

The S-matrix elements contain various physical channels in which tachyon, massless, and an infinite tower of massive states propagate. To extract spacetime field-theory couplings from these S-matrix elements, one must expand the massive poles. However, since we are interested in a field theory that includes both the tachyon and the massless fields, one should not, in general, expand the tachyon poles. It has been argued in \cite{Garousi:2026nvz} that consistency with general covariance requires the effective action to exclude couplings with an odd number of tachyons, even though the S-matrix elements for an odd number of tachyons are nonzero. The expansion should be such that the massless or tachyon poles produced by the kinetic terms of the tachyon and massless fields are reproduced by the leading order of the S-matrix expansion \cite{Garousi:2002wq,Garousi:2003ur,Bitaghsir-Fadafan:2006iya,Garousi:2003pv,Garousi:2003db,Garousi:2026nvz}. Unlike the expansion of S-matrix elements for massless vertex operators, the expansion of S-matrix elements involving tachyon and massless vertex operators is not a pure momentum expansion. That is, at each order of expansion, not all terms are of the same order in the external momenta. On the other hand, after using on-shell relations, the field-theory couplings at each derivative order are also not all of the same order in momenta. It is expected that the expansion of S-matrix elements corresponds to the tachyon potential and the higher-derivative expansion of the tachyon and massless fields in the field theory \cite{Garousi:2002wq,Garousi:2003ur,Bitaghsir-Fadafan:2006iya,Garousi:2003pv,Garousi:2003db,Garousi:2026nvz}.

The S-matrix elements are defined for external on-shell states, so there is an ambiguity in extracting field-theory couplings from string-theory S-matrix elements \cite{Tseytlin:1991bu,Banks:1991sg}. For example, the couplings \(R T \partial_\mu \partial^\mu T\) and \(-R T^2\) are equivalent on-shell. To resolve this ambiguity, we use T-duality to fix the couplings and rely on comparison with the S-matrix only for parameters that remain unfixed by T-duality. The T-duality method has been successfully used to derive the classical effective action of massless NS-NS fields in bosonic \cite{Garousi:2019wgz,Garousi:2019mca,Wulff:2024ips,Ameri:2025bei}, heterotic \cite{Garousi:2019wgz,Garousi:2023kxw,Pahlavan:2026drq}, and superstring theories \cite{Garousi:2020gio,Garousi:2020lof,Garousi:2022ghs} up to order \(\alpha'^3\). It has also been applied to determine Yang–Mills couplings at order \(\alpha'\) and \(\alpha'^2\) in the heterotic theory \cite{Garousi:2024avb,Garousi:2024imy}. In this paper, we extend this method to include both the tachyon and massless fields at four-derivative order in the classical effective action of bosonic string theory. However, since the tachyon is invariant under T-duality, this method is not as powerful as for massless fields, for which T-duality fully fixes the couplings up to an overall factor. The remaining parameters, such as those in the tachyon potential, are fixed by  the corresponding string-theory S-matrix elements.

To apply the T-duality method, one requires a minimal basis of covariant couplings with arbitrary coupling constants. Following the observation in \cite{Garousi:2026nvz}, the basis should include only an even number of tachyon fields. We find that, due to the presence of the tachyon potential, there exists a scheme in which all couplings involve only derivatives of the tachyon and the other massless fields, while the tachyon without derivatives appears only in the tachyon potential. This immediately resolves the on-shell ambiguity of the effective action discussed in \cite{Tseytlin:1991bu,Banks:1991sg}. Within this general scheme, we construct the minimal basis and then impose T-duality to establish relations among the parameters. The remaining unfixed parameters are subsequently determined by comparison with S-matrix elements.

The structure of the paper is as follows. In Section 2, we show that there exists a general scheme in which the tachyon without derivatives appears only in the tachyon potential, while all other terms involve derivatives of the tachyon. We then write down the leading-order action, which includes the tachyon potential, the kinetic term of the tachyon, and the universal NS-NS couplings at two-derivative order. This action is manifestly invariant under T-duality. In Section 3, we extend the analysis to four-derivative order. In Subsection 3.1, we construct the minimal basis, which consists of 14 independent terms with arbitrary coupling constants. In Subsection 3.2, we impose T-duality invariance, which fixes 8 of the coupling constants to zero and expresses the remaining 6 couplings in terms of two parameters. One of these is fixed by comparing with the zero-tachyon case. To determine the last parameter via the S-matrix, we use field redefinitions in Subsection 3.3 to rewrite the couplings in a scheme where the propagators of the leading-order action are not modified by the four-derivative couplings. In Section 4, we employ the S-matrix method to fix the remaining parameter and to confirm consistency with the expansion of S-matrix elements. In Subsection 4.1, we compare the couplings with the expansion of the S-matrix element of four tachyon vertex operators. This fixes the remaining parameter in the four-derivative effective action. The same expansion also contains terms without momentum, which fix the coefficient of the \(T^4\) term in the tachyon potential. Remarkably, this coefficient is found to be positive, indicating the existence of a minimum in the tachyon potential. In Subsection 4.2, we compare the expansion of the S-matrix element of two tachyons and two graviton vertex operators and show that the four-momentum part of this expansion is fully consistent with the four-derivative effective action obtained from T-duality. A brief discussion of our results and their implications is provided in Section 5.

\section{Leading order effective action}

Recently, it has been argued, based on the expansion of S-matrix elements, that the covariant effective action of bosonic string theory should include only an even number of closed string tachyons \cite{Garousi:2026nvz}. It should also be invariant under T-duality. Since the tachyon is invariant under T-duality, the most general effective action may take the following form:
 \beqa
 \bS_{\rm leading} &=& \frac{2}{\kappa^2}\int d^{26}x \, \sqrt{-G} \, e^{-2\Phi} \Big[(1+ f(T))\left( R + 4\partial_\mu \Phi \partial^\mu \Phi - \frac{1}{12}H^2\right)\nn\\&&\qquad\qquad\qquad\qquad\qquad  - \frac{1}{4}(1+g(T))\partial_\mu T \partial^\mu T - V(T) \Big]\,, \labell{Sbulk1}
 \eeqa
where \(f(T)\), \(g(T)\), and \(V(T)\) are even functions of the tachyon, beginning at order \(T^2\). The tachyon potential is expanded as
\beqa
V(T)&=&-\frac{1}{2}T^2+c_1T^4+c_2T^6+\cdots,\labell{V}
\eeqa
where the coefficients \(c_1,c_2,\ldots\) are parameters to be fixed by S-matrix computations. The coefficient of the quadratic term is fixed by the fact that the bosonic string tachyon has mass \(m^2=-2\), with our convention \(\alpha'=2\). The tachyon potential contributes at zeroth order in derivatives, while all other terms in the above action are at two-derivative order.

Under a higher-derivative field redefinition of the form
\[
T \rightarrow T +\delta T,
\]
where \(\delta T\) contains arbitrary combinations of massless and tachyon fields with an odd number of tachyons, the tachyon potential generates the following contribution:
\[
\delta \!\!\bS = -\frac{2}{\kappa^2}\int d^{26}x \, \sqrt{-G} \, e^{-2\Phi}\left[\frac{dV}{dT}\delta T\right].
\]
Since the derivative of the tachyon potential contains an odd number of tachyons, this contribution involves \(2,4,6,\ldots\) tachyon fields. By choosing an appropriate nonsingular \(\delta T\) at order \(\alpha'\), one can absorb all two-derivative couplings in the action  \reef{Sbulk1} that have coefficients \(f(T)\) and \(g(T)\). Moreover, by choosing \(\delta T\) at higher orders in \(\alpha'\), one can absorb all higher-derivative couplings of the form like \(\alpha'^n T^{2k} R^{n+1}\) for \(k=1,2,\ldots\).

One may also consider higher-derivative field redefinitions of the metric and dilaton. Such redefinitions act on the tachyon potential and produce terms of the form
\[
\delta \!\!\bS = -\frac{2}{\kappa^2}\int d^{26}x \, \sqrt{-G} \, e^{-2\Phi}\left[\frac{1}{2}\delta G^\mu{}_\mu - 2\delta\Phi\right]V(T).
\]
Using ordinary field redefinitions (those without tachyons) generates higher-derivative couplings proportional to \(V(T)\). However, all such couplings can be absorbed into \(\delta T\). Therefore, by choosing an appropriate \(\delta T\), one finds a general scheme in which all couplings involve only derivatives of the tachyon and the massless fields, while the tachyon itself—without derivatives—appears exclusively in the tachyon potential.

In this scheme, the leading-order action \reef{Sbulk1} reduces to
\beqa
\bS_{\rm leading}& =& \frac{2}{\kappa^2}\int d^{26}x \, \sqrt{-G} \, e^{-2\Phi} \Big[ R + 4\partial_\mu \Phi \partial^\mu \Phi - \frac{1}{12}H^2 - \frac{1}{4}\partial_\mu T \partial^\mu T -V(T) \Big]\,, \labell{Sbulk2}
\eeqa
and all higher-derivative couplings involve only covariant derivatives of the tachyon and the massless fields. It is important to note that in this general scheme, one is not allowed to use higher-derivative field redefinitions of the tachyon, as they would take the action out of the scheme. However, higher-derivative redefinitions of the metric, dilaton, and $B$-field that involve covariant derivatives of the tachyon and massless fields are allowed and preserve the scheme.

The above leading-order action is invariant under T-duality, and all higher-order terms should also respect this symmetry. The full effective action beyond leading order can therefore be organized as an expansion in derivatives of the tachyon and massless fields to all orders, taking the form
\beqa
\bS=\sum_{n=1}^{\infty}\alpha'^n\bS^{(n)}&;& \bS^{(n)}=\frac{2}{\kappa^2}\int d^{26}x\sqrt{-G}\,e^{-2\Phi}\cL^{(n)}\,,\labell{Sn}
\eeqa 
where at each order \(n\), \(\mathcal{L}^{(n)}\) consists of a minimal basis of gauge-invariant couplings involving the tachyon and massless fields, with unknown coefficients to be determined by imposing T-duality and by comparison with S-matrix elements. In the next section, we apply T-duality constraints within this scheme to determine the four-derivative couplings.

\section{Four-derivative effective action}

It is known that the classical effective action of bosonic string theory must be invariant under T-duality at all orders of $\alpha'$ \cite{Sen:1991zi,Hohm:2014sxa}. Specifically, when the couplings are reduced on a circle, the resulting couplings in the base space should remain invariant under the Buscher rules, supplemented by higher-derivative corrections \cite{Kaloper:1997ux,Garousi:2019wgz}. To impose this constraint in order to determine the effective action, one must first construct a minimal basis of independent couplings at four-derivative order, and then impose T-duality to fix the free parameters of this basis.

\subsection{Minimal basis}

The prescription for constructing the minimal basis is given in \cite{Metsaev:1987zx,Garousi:2019cdn}. Following that references, one must consider all scalar contractions of \(R_{\mu\nu\alpha\beta}\), \(H_{\mu\nu\alpha}\), \(\nabla_\mu\Phi\), \(\nabla_\mu T\), and their derivatives at order \(\alpha'\). A priori, there are 56 such terms. We then incorporate the contributions from arbitrary field redefinitions of the metric, \(B\)-field, and dilaton, add arbitrary total derivative terms, and impose the appropriate Bianchi identities to identify the independent couplings. Following this procedure, we find 14 independent couplings. In a particular scheme, these couplings take the following form:
\beqa
\cL^{(1)}&=&a_{1}  H_{\alpha  }{}^{\delta  \epsilon  } H^{\alpha  
\beta  \gamma  } H_{\beta  \delta  }{}^{\varepsilon  } 
H_{\gamma  \epsilon  \varepsilon  } +  
 a_{2}  H_{\alpha  \beta  }{}^{\delta  } H^{\alpha  
\beta  \gamma  } H_{\gamma  }{}^{\epsilon  \varepsilon  } 
H_{\delta  \epsilon  \varepsilon  } +  
 a_{3}  H_{\alpha  \beta  \gamma  } H^{\alpha  \beta  
\gamma  } H_{\delta  \epsilon  \varepsilon  } H^{\delta  
\epsilon  \varepsilon  } \nn\\&&+  
 a_{4} R_{\alpha  \beta  \gamma  \delta  } 
R^{\alpha  \beta  \gamma  \delta  } +  
 a_{5}  H_{\alpha  }{}^{\delta  \epsilon  } H^{\alpha  
\beta  \gamma  }R_{\beta  \gamma  \delta  \epsilon  } 
+  
  a_{6}  H_{\beta  \gamma  \delta  } H^{\beta  \gamma  
\delta  } \nabla_{\alpha  }T \nabla^{\alpha  }T\nn\\&& +  
 a_{7}  H_{\beta  \gamma  \delta  } H^{\beta  \gamma  
\delta  } \nabla_{\alpha  }\Phi  \nabla^{\alpha  }\Phi  +  
 a_{8}  \nabla_{\alpha  }\nabla^{\alpha  }T 
\nabla_{\beta  }\nabla^{\beta  }T +  
 a_{9}  H_{\alpha  }{}^{\gamma  \delta  } H_{\beta  
\gamma  \delta  } \nabla^{\alpha  }T \nabla^{\beta  }T\nn\\&& +  
 a_{10}  \nabla_{\alpha  }T \nabla^{\alpha  }T \nabla_{
\beta  }T \nabla^{\beta  }T +  
 a_{11}  \nabla_{\alpha  }\Phi  \nabla^{\alpha  }T 
\nabla_{\beta  }\Phi  \nabla^{\beta  }T +  
 a_{12}  H_{\alpha  }{}^{\gamma  \delta  } H_{\beta  
\gamma  \delta  } \nabla^{\alpha  }\Phi  \nabla^{\beta  }\Phi\nn\\&& 
 +  
  a_{13}  \nabla_{\alpha  }T \nabla^{\alpha  }T 
\nabla_{\beta  }\Phi  \nabla^{\beta  }\Phi  +  
 a_{14}  \nabla_{\alpha  }\Phi  \nabla^{\alpha  }\Phi  
\nabla_{\beta  }\Phi  \nabla^{\beta  }\Phi\,,\labell{L1}
\eeqa
where \(a_1,\ldots,a_{14}\) are coupling constants to be determined by imposing T-duality and by comparison with S-matrix elements. Note that in the limit where the tachyon vanishes, the above basis reduces to the 8 independent couplings, which constitute the known minimal basis for the metric, \(B\)-field, and dilaton at order \(\alpha'\) \cite{Metsaev:1987zx}. Since the tachyon is invariant under T-duality, the term proportional to \(a_{10}\) is automatically T-duality invariant. All other terms, however, must be related to one another under T-duality transformations.

 \subsection{T-duality constraint}
 
The T-duality constraint at order \(\alpha'\) is given by \cite{Garousi:2019wgz} 
\beqa  
S^{(0,1)}_{(1)}(\psi_0) \sim S^{(1)}(\psi) - S^{(1)}(\psi_0)\,, \labell{T1}  
\eeqa 
where the first term on the right-hand side is the reduction of \(\!\bS^{(1)}\!\) on a circle, which is performed using the following circular reduction ansatz for the basic fields \cite{Maharana:1992my}:
\beqa  
G_{\mu\nu} = \left(\matrix{\bg_{ab} + e^{\varphi} g_{a} g_{b} &  e^{\varphi} g_{a} \cr e^{\varphi} g_{b} & e^{\varphi} &}\!\!\!\!\!\right), \quad  
B_{\mu\nu} = \left(\matrix{\bb_{ab} + b_{[a} g_{b]} & b_{a} \cr -b_{b} & 0 &}\!\!\!\!\!\right), \quad  
\Phi = \bar{\phi} + \varphi/4,  \quad T=\bar{T}\,,
\labell{reduction}  
\eeqa  
where indices \(a,b\) denote directions orthogonal to the Killing coordinate \(y\). In the above reduction, \(\bar{g}_{ab}\) is the \((D-1)\)-dimensional base-space metric, \(\bar{b}_{ab}\) is an antisymmetric tensor, \(\bar{\phi}\) is the dilaton, \(\bar{T}\) is the tachyon, \(\varphi\) is a scalar, and \(g_a, b_a\) are two vectors.

The second term on the right-hand side of \reef{T1} is the transformation of \(S^{(1)}\) under the Buscher rules \cite{Buscher:1987sk,Rocek:1991ps}, which for the above reduction ansatz take the form
\beqa  
&&\bg_{ab}' = \bg_{ab}\,, \quad 
\bar{b}_{ab}' = \bar{b}_{ab}\,, \quad 
\bar{\phi}' = \bar{\phi}\,, \nn\\
&&\varphi' = -\varphi\,, \quad 
g_a' = b_a\,, \quad 
b_a' = g_a\,, \quad \bar{T}'=\bar{T}\,. 
\labell{flac}  
\eeqa  
The term on the left-hand side of \reef{T1} is the transformation of the Taylor expansion of the reduced two-derivative action under the Buscher rules. This two-derivative action is given by
\beqa
S^{(0)}(\psi)&=&\frac{2}{\kappa^2}\int\mathrm{d}^{25}x\sqrt{-\bg}e^{-2\bphi}\big[\bar{R}-\nabla_a\nabla^a\vp+4\nabla_a\bphi\nabla^a\bphi+2\nabla_a\vp\nabla^a\bphi-\frac{1}{4}\nabla_a\vp\nabla^a\vp\nn\\&&-\frac{1}{4}e^\vp V_{ab}V^{ab}-\frac{1}{4}e^{-\vp}W_{ab}W^{ab}-\frac{1}{12}\bH_{abc}\bH^{abc}-\frac{1}{4}\nabla_a\bar{T}\nabla^a\bar{T}\Big].\labell{S00}
\eeqa
This left-hand side arises because the T-duality transformation is given by the Buscher rules \reef{flac}  supplemented by higher-derivative corrections, each with its own parameter. The Taylor expansion of the leading-order action under these transformations then produces contributions at order \(\alpha'\). The symbol \(\sim\) in \reef{T1} indicates that the two sides are equal up to total derivative terms in the base space and up to the use of appropriate Bianchi identities. The T-duality constraint \reef{T1} establishes relations among the coupling constants in the minimal basis \reef{L1}, as well as between the parameters of the Buscher-rule corrections and the coupling constants. Importantly, the relations among the coupling constants derived from the T-duality constraint \reef{T1} are independent of the base-space geometry \cite{Garousi:2019mca}. This allows us to considerably simplify the calculation by assuming a flat base space, which we adopt in the present work.

It is important to note that the most general corrections to the Buscher rules, when applied to the tachyon potential in the base space, generate contributions of the form
\beqa
\delta S&=&-\frac{2}{\kappa^2}\int d^{25}x \, \sqrt{-\bg} \, e^{-2\bphi}\Big[\left(\frac{1}{2}\delta \bg^a{}_a-2\delta\bphi\right)V(\bar{T})+\frac{dV}{d\bar{T}}\delta\bar{T}\Big]\,.\labell{dV}
\eeqa
These terms produce higher-derivative couplings in the base space that contain tachyon fields without derivatives. Since the higher-derivative effective action in our scheme contains no such terms (i.e., no tachyons without derivatives), the corrections to the Buscher rules must be chosen such that the above contribution vanishes. This requires
\beqa
\delta \bar{g}^a{}_a - 4\delta \bar{\phi} = 0,\labell{dV1}
\eeqa
and that there be no higher-derivative corrections to the transformation \(\bar{T}' = \bar{T}\), i.e., \(\delta \bar{T} = 0\).

Imposing the T-duality constraint \reef{T1} then fixes the minimal basis up to two parameters. One of these is determined by comparing with the zero-tachyon case, while the other remains unfixed by T-duality alone. Our result is the following:
 \beqa
{\bf S}^{(1)}_{\rm MT}&=&\frac{2}{4\kappa^2}\int d^{26}x \sqrt{-G}e^{-2\Phi}\Big[R_{\alpha \beta \gamma \delta } R^{\alpha \beta \gamma \delta }+ \frac{1}{24} H_{\alpha }{}^{\delta \epsilon } H^{\alpha \beta \gamma } H_{\beta \delta }{}^{\varepsilon } H_{\gamma \epsilon \varepsilon } \nn\\&&\qquad\qquad\qquad\qquad\qquad  -  \frac{1}{8} H_{\alpha \beta }{}^{\delta } H^{\alpha \beta \gamma } H_{\gamma }{}^{\epsilon \varepsilon } H_{\delta \epsilon \varepsilon } -  \frac{1}{2} H_{\alpha }{}^{\delta \epsilon } H^{\alpha \beta \gamma } R_{\beta \gamma \delta \epsilon }\nn\\&&\qquad\qquad\qquad\qquad\qquad
-\frac{1}{4}H_\alpha{}^{\gamma\delta}H_{\beta\gamma\delta}\nabla^\alpha T\nabla^\beta T+a(\nabla_\alpha T\nabla^\alpha T)^2\Big]\,.\labell{fourmin}
\eeqa
The corresponding corrections to the Buscher rules are exactly those found in \cite{Garousi:2019wgz}. In particular, we find no tachyon corrections to the Buscher rules. In the zero-tachyon limit, the above action reduces to the effective action in the Metsaev–Tseytlin scheme \cite{Metsaev:1987zx}. The parameter $a$ is a constant that cannot be fixed by T-duality and must therefore be determined by comparison with S-matrix elements.

It is important to note that the terms involving tachyons in the last line of \reef{fourmin} are subject to change under field redefinitions. Moreover, in order to compare the above couplings with S-matrix results, one must perform field redefinitions to express the Riemann-squared term in the Gauss–Bonnet combination, which does not modify the graviton propagator. We carry out this procedure in the next subsection.

\subsection{Effective action in Meissner  scheme}

Using field redefinitions, one can transform the action in \reef{fourmin} into various equivalent forms. We employ field redefinitions to bring the tachyon-independent terms into the Meissner scheme \cite{Meissner:1996sa}, in which the propagators of the leading-order action receive no higher-derivative corrections. These field redefinitions remove the couplings between the tachyon and the \(B\)-field, and also modify the coefficient of the four-tachyon coupling in \reef{fourmin}. Our result is the following:
\beqa
{\bf S}_{\mathrm{M}}^{(1)}&=&\frac{2}{4\kappa^2}\int d^{26}x \sqrt{-G}e^{-2\Phi}\Big[ \frac{1}{24} H_{\alpha 
}{}^{\delta \epsilon } H^{\alpha \beta \gamma } H_{\beta 
\delta }{}^{\varepsilon } H_{\gamma \epsilon \varepsilon } -  
\frac{1}{8} H_{\alpha \beta }{}^{\delta } H^{\alpha \beta 
\gamma } H_{\gamma }{}^{\epsilon \varepsilon } H_{\delta 
\epsilon \varepsilon } \nn\\&&+ \frac{1}{144} H_{\alpha \beta \gamma 
} H^{\alpha \beta \gamma } H_{\delta \epsilon \varepsilon } H^{
\delta \epsilon \varepsilon } + H_{\alpha }{}^{\gamma \delta } 
H_{\beta \gamma \delta } R^{\alpha \beta } - 4 
R_{\alpha \beta } R^{\alpha \beta } -  
\frac{1}{6} H_{\alpha \beta \gamma } H^{\alpha \beta \gamma } 
R + R^2 \nn\\&&+ R_{\alpha \beta \gamma 
\delta } R^{\alpha \beta \gamma \delta } -  
\frac{1}{2} H_{\alpha }{}^{\delta \epsilon } H^{\alpha \beta 
\gamma } R_{\beta \gamma \delta \epsilon } -  
\frac{2}{3} H_{\beta \gamma \delta } H^{\beta \gamma \delta } 
\nabla_{\alpha }\nabla^{\alpha }\Phi + \frac{2}{3} H_{\beta 
\gamma \delta } H^{\beta \gamma \delta } \nabla_{\alpha }\Phi 
\nabla^{\alpha }\Phi \nn\\&&+ 8 R \nabla_{\alpha }\Phi 
\nabla^{\alpha }\Phi - 16 R_{\alpha \beta } 
\nabla^{\alpha }\Phi \nabla^{\beta }\Phi + 16 \nabla_{\alpha 
}\Phi \nabla^{\alpha }\Phi \nabla_{\beta }\Phi \nabla^{\beta 
}\Phi - 32 \nabla^{\alpha }\Phi \nabla_{\beta }\nabla_{\alpha 
}\Phi \nabla^{\beta }\Phi \nn\\&&+ 2 H_{\alpha }{}^{\gamma \delta } 
H_{\beta \gamma \delta } \nabla^{\beta }\nabla^{\alpha }\Phi+(\frac{3}{16}+ a)(\nabla_\alpha T\nabla^\alpha T)^2\Big]\,.\labell{fourmax}
\eeqa
The above couplings can now be compared with string theory S-matrix elements. A key observation is that there should be no contact terms at order \(\alpha'\) in the sphere-level S-matrix element involving two tachyons and two massless NS-NS vertex operators. Moreover, the S-matrix element of four tachyon vertex operators should fix both the parameter \(a\) in the above four-derivative couplings and the parameter \(c_1\) in the tachyon potential \reef{V}. In the next section, we carry out this analysis.

\section{Comparing with S-matrix}

In this section, we compare the effective action \(\!\bS^{(1)}_{\rm M}\!\) in \reef{fourmax}, obtained from T-duality, with the expansion of the corresponding sphere-level S-matrix elements in bosonic string theory. This comparison serves to fix the parameter \(a\) in the action, as well as the parameter \(c_1\) in the tachyon potential \reef{V}. It should also confirm the absence of four-derivative couplings between two tachyons and two massless NS-NS fields.

The evaluation of four-point S-matrix elements involving tachyon vertex operators in bosonic string theory is straightforward. However, the expansion of these S-matrix elements is not, because each physical channel contains tachyon and massless poles, as well as an infinite tower of massive poles. Recently, it has been proposed that, in each channel, either the massless or the tachyon poles, along with all massive poles, must be expanded to generate higher-momentum contact terms  \cite{Garousi:2026nvz}. The remaining massless or tachyon pole, together with the higher-momentum contact terms, must then be reproduced by field theory. The choice of whether to expand the massless or the tachyon pole depends on whether that pole is reproduced by a field theory that involves only an even number of tachyons. If the pole is reproduced by the field theory, it should be kept; if not, it must be expanded. The expanded terms are then reproduced by higher-derivative terms in the same field theory, which contains only even numbers of tachyons. It has been argued in  \cite{Garousi:2026nvz} that couplings with an odd number of tachyons are inconsistent with a diffeomorphism-invariant effective action and also incompatible with T-duality. In the next subsection, we analyze the S-matrix element of four tachyons.

\subsection{Four-tachyon amplitude}

The world-sheet S-matrix element of four tachyon vertex operators is the well-known Virasoro–Shapiro amplitude \cite{  Virasoro:1969me,Shapiro:1969km}
\beqa
A&=&\alpha\frac{\Gamma(-1-s/2)\Gamma(-1-t/2)\Gamma(-1-u/2)}{\Gamma(2+s/2)\Gamma(2+t/2)\Gamma(2+u/2)}\,,
\eeqa
where \(\alpha\) is a normalization constant to be fixed by comparison with field theory, and the Mandelstam variables are defined as
\beqa
s=-(p_1+p_2)^2\,,\,\,t=-(p_1+p_4)^2\,,\,\,u=-(p_1+p_3)^2\,.
\eeqa
The momenta satisfy the on-shell relation \(p_i^2=2\) for \(i=1,2,3,4\). The amplitude is symmetric under exchange of \(s,t,u\). Momentum conservation implies that the Mandelstam variables satisfy the on-shell relation
\beqa
s+t+u=-8\,.\labell{stu4T}
\eeqa

To expand the amplitude, we note that the two-derivative effective action \reef{Sbulk2} produces massless poles in the \(s\)-, \(t\)-, and \(u\)-channels. However, because of the on-shell relation  \reef{stu4T}, we cannot send all three variables to zero simultaneously. One must instead send \(s \to 0\) with \(t,u \to -4\), or \(t \to 0\) with \(s,u \to -4\), or \(u \to 0\) with \(s,t \to -4\). It is therefore convenient to define new Mandelstam variables 
 \beqa
  s'\,=\,s+4=-2p_1\cdot p_2\,, \,\,\,t'\,=\,t+4=-2p_1\cdot p_4\,, \,\,\,u'\,=\,u+4=-2p_1\cdot p_3\,,
 \eeqa
and rewrite the amplitude as 
 \beqa
A&=&-(\alpha/3)\Big[\frac{1}{(1+s/2)^2}\frac{\Gamma(1-t'/2)\Gamma(-s/2)\Gamma(1-u'/2)}{\Gamma(t'/2)\Gamma(1+s/2)\Gamma(u'/2)}\nn\\&&\qquad\quad+\frac{1}{(1+t/2)^2}\frac{\Gamma(1-s'/2)\Gamma(-t/2)\Gamma(1-u'/2)}{\Gamma(s'/2)\Gamma(1+t/2)\Gamma(u'/2)}\nn\\&&\qquad\quad+\frac{1}{(1+u/2)^2}\frac{\Gamma(1-s'/2)\Gamma(-u/2)\Gamma(1-t'/2)}{\Gamma(s'/2)\Gamma(1+u/2)\Gamma(t'/2)}\Big]\,.\labell{Astring}
\eeqa
The first, second, and third lines in \reef{Astring} contain tachyon and massless poles in the \(s\)-, \(t\)-, and \(u\)-channels, respectively. The expansion is then performed in the limit \(s,t,u,s',t',u' \to 0\), keeping the massless poles and expanding the tachyon poles.

The expansion of the tachyon pole in the first line is
\beqa
\frac{1}{(1+s/2)^2}&=&1-s+\frac{3}{4}s^2+\cdots\,,\labell{exT}
\eeqa
and similarly for the other tachyon poles. The expansion of the Gamma functions in the first line is
\beqa
\frac{\Gamma(1-t'/2)\Gamma(-s/2)\Gamma(1-u'/2)}{\Gamma(t'/2)\Gamma(1+s/2)\Gamma(u'/2)}&=&-\frac{u't'}{2s}+\frac{u't'}{8}(su'+u'^2)\z(3)+\cdots\,,\labell{exG}
\eeqa
with analogous expressions for the second and third lines. The massless pole in the \(s\)-channel is therefore
\beqa
A_s&=&\frac{\alpha}{6}\frac{u't'}{s}\,,
\eeqa
with similar poles in the \(t\)- and \(u\)-channels. All other terms in \reef{Astring} are contact terms that should be reproduced by the higher-derivative effective action.

The \(s\)-channel massless pole in the  two-derivative field theory \reef{Sbulk2}  has been computed in \cite{Garousi:2003db} as \footnote{ Note that in \cite{Garousi:2003db}, the metric perturbation is defined as \(g_{\mu\nu} + h_{\mu\nu}\), while in the present work we adopt \(g_{\mu\nu} + \kappa h_{\mu\nu}\) and set \(T = \kappa \tau\).}
\beqa
A'_s&=&-\frac{i\kappa^2}{16s}((u-t)^2-s^2)\,=\,\frac{i\kappa^2}{4}\frac{u't'}{s}\,.
\eeqa
Thus, if we fix the string amplitude normalization to \(\alpha = 3i\kappa^2/2\), we obtain exact agreement for the massless pole.

We now turn to the contact terms. There are no contact terms at two-derivative order, which is consistent with the fact that the effective action \reef{Sbulk2}  contains no coupling of the form \(T^2 \partial T \partial T\). The contact terms quadratic in the Mandelstam variables are
\beqa
A^{(2)}_c&=&-\frac{\alpha}{6}(u't'+u's'+t's')\nn\\&=&-\frac{\alpha}{6}\Big[8-2(p_1\cdot p_3)^2-2( p_1\cdot p_4)^2-2(p_1\cdot p_2 )^2\Big]\,,
\eeqa
where in the second line we have used the on-shell relations. Note that this contact term arises because the amplitude in bosonic string theory has a tachyon pole. For type 0 theory, which has no tachyon pole in the corresponding amplitude, there is no such contact term \cite{Garousi:2003db}. The field-theory contact term at zero momentum is \(-2 \times 4! \, c_1 i\kappa^2\), which yields \(c_1 = 1/24\). The tachyon potential therefore becomes
\beqa
V(T)&=&-\frac{1}{2!}T^2+\frac{1}{4!}T^4+\cdots\,.\labell{VT}
\eeqa
It is interesting to note that the coefficient of the \(T^4\) term is positive. This suggests that even though bosonic string theory is unstable at \(T=0\), the potential may have a stable minimum after tachyon condensation. However, determining the minimum requires knowledge of the \(T^6\) and higher-order terms, which we do not pursue in this paper.

The field-theory action \reef{fourmax}  has a contact term at fourth order in momentum, which is
\beqa
A'_c&=&4i\kappa^2(\frac{3}{16}+a)\Big[(p_1\cdot p_3)^2+( p_1\cdot p_4)^2+(p_1\cdot p_2 )^2\Big]\,,
\eeqa
which fixes the parameter \(a\) of the tachyon coupling in the Metsaev–Tseytlin scheme to \(a = -1/16\). Correspondingly, the coupling constant of the tachyon couplings in the Meissner scheme becomes \(1/8\).

One might ask whether the contact terms at higher orders in the Mandelstam variables contribute to the coefficient of \(T^4\) or to the coefficient of \((\partial_\mu T \partial^\mu T)^2\). If so, the values of \(c_1\) and \(a\) determined above would be modified upon inclusion of higher-derivative couplings. While we cannot currently prove that this is not the case, the contact terms cubic in the Mandelstam variables indicate that no such contribution exists. To see this, consider the contact terms at third order:
 \beqa
A^{(3)}_c&\!\!\!\!=\!\!\!\!&\frac{\alpha}{8}(s u't'+t u's'+u t's')\\&\!\!\!\!=\!\!\!\!&\frac{\alpha}{8}\Big[-16(p_1\cdot p_3)^2-16( p_1\cdot p_4)^2-16(p_1\cdot p_2 )^2-8(p_1\cdot p_3)^3-8( p_1\cdot p_4)^3-8(p_1\cdot p_2 )^3\Big],\nn
\eeqa
where in the second line we have used the on-shell relations. Using these relations, one can verify that six-momentum terms are required for the six-derivative coupling \(\partial_a \partial_b T \partial^a \partial^b T \partial_c T \partial^c T\), while both six-momentum and four-momentum terms are required for the six-derivative coupling \(\partial_a \partial_b T \partial^a T \partial_c T \partial^b \partial^c T\).
We expect that contact terms at \(4,5,\ldots\) orders in the Mandelstam variables are similarly related to \(8,10,\ldots\) derivative terms in the effective action. We do not pursue the six-derivative or higher-order couplings further in this paper. It is important, however, to emphasize that the six-derivative couplings produce six- and four-momentum terms after using the on-shell relations. Similar structures are expected to arise for higher-derivative terms as well. In the next subsection, we show that the momentum-expansion contact terms of the S-matrix element involving two tachyons and two gravitons at four-derivative order are also fully consistent with the effective action \reef{fourmax}, which produces no such contact terms.

\subsection{Two-tachyon two-graviton amplitude}

The scattering amplitude of two tachyons and two gravitons is given by the following world-sheet correlation function:
\beqa
A&=&\beta<Q_1Q_2Q_3Q_4>\,,\labell{QQQQ}
\eeqa
where \(\beta\) is a normalization constant to be fixed by comparison with field theory. The tachyon vertex operator is
\beqa
Q_i&=&\int d^2z_i:e^{ip_i\cdot X(z_i)}:\,:e^{ip_i\cdot \bX(\bz_i)}:\,,
\eeqa
for \(i=1,2\), with the on-shell relation \(p_i^2 = 2\). The graviton vertex operator is
\beqa
 Q_j &=& (\epsilon_j )_{\mu\nu} \int d^2z_j \, :\partial_{z_j} X^\mu(z_j) e^{ip_j\cdot X(z_j)}:\,:\partial_{\bar{z}_j} X^\nu(\bar{z}_j) e^{ip_j\cdot X(\bar{z}_j)}:\,,
 \eeqa
for \(j=3,4\), with the on-shell relations \(p_j^2 = 0\) and \(p_j \cdot \epsilon_j = 0\). The graviton polarization \(\epsilon_j^{\mu\nu}\) is symmetric. The same vertex operator applies to the \(B\)-field, for which the polarization is antisymmetric, and to the dilaton, for which the polarization is
\beqa
 \epsilon_j^{\mu\nu} &=& \frac{1}{\sqrt{D-2}} \left( \eta^{\mu\nu} - p_j^\mu \ell_j^\nu - p_j^\nu \ell_j^\mu \right)\,,\nn
 \eeqa
where \(D\) is the spacetime dimension, and the auxiliary vector \(\ell_j^\mu\) satisfies \(p_j \cdot \ell_j = 1\). In this paper, we consider only the graviton polarization.

The world-sheet propagators for the holomorphic field \(X(z)\) and the antiholomorphic field \(\bar{X}(\bar{z})\) are given by
\beqa
 \langle X^{\mu}(z) X^{\nu}(w) \rangle \,=\, -\eta^{\mu\nu} \log(z-w)\,&;&\langle \bX^{\mu}(\bz) \bX^{\nu}(\bw) \rangle \,=\, -\eta^{\mu\nu} \log(\bz-\bw)\,.\nn
 \eeqa
Using these propagators, one can evaluate the correlator in \reef{QQQQ} and show that the resulting integrand is invariant under \(SL(2,\mathbb{C})\). Fixing this symmetry by setting \(z_4 = \infty\), \(z_3 = 0\), and \(z_2 = 0\), one finds
\beqa
 A&=&\beta\Tr(\epsilon_3\cdot\epsilon_4)\int d^2z_1|z_1|^{2p_1\cdot p_3}|1-z_1|^{2p_1\cdot p_2}+\cdots\,,
 \eeqa
where the dots represent terms in which the graviton polarizations contract with momenta. For simplicity, we do not consider such terms. The integral can be expressed in terms of Gamma functions, as in the four-tachyon amplitude \cite{  Virasoro:1969me,Shapiro:1969km}. In terms of Mandelstam variables, the amplitude can be written as (see \eg Appendix in \cite{Garousi:2003db})
\beqa
A&=&2\pi\beta\Tr(\epsilon_3\cdot\epsilon_4)\frac{\Gamma(-u/2)\Gamma(-1-s/2)\Gamma(-t/2)}{\Gamma(1+u/2)\Gamma(2+s/2)\Gamma(1+t/2)}\,.
\eeqa
The Mandelstam variables in this case satisfy
\beqa
s+t+u=-4\,.\labell{stu2T2G}
\eeqa

We now expand this amplitude so that it produces the poles generated by the leading-order action \reef{Sbulk2}. Since we consider only the term in which the two graviton polarizations contract with each other, field theory produces only a massless pole in the \(s\)-channel. The expansion is therefore taken at \(s \to 0\). To satisfy the on-shell relation \reef{stu2T2G}, the other two Mandelstam variables must be sent to \(t,u \to -2\). It is therefore convenient to define new Mandelstam variables:
\beqa
t'=t+2=-2p_1\cdot p_4\,,\,\, u'=u+2=-2p_1\cdot p_3\,.
\eeqa
The amplitude can then be written as
\beqa
A&=&-2\pi\beta\Tr(\epsilon_3\cdot\epsilon_4)\frac{1}{(1+s/2)^2}\frac{\Gamma(1-u'/2)\Gamma(-s/2)\Gamma(1-t'/2)}{\Gamma(u'/2)\Gamma(1+s/2)\Gamma(t'/2)}\,.\labell{ATG}
\eeqa
The amplitude contains tachyon and massless poles in the \(s\)-channel. The expansion is then performed in the limit \(s,t',u' \to 0\), keeping the massless pole and expanding the tachyon pole.

Using the expansions \reef{exT} and \reef{exG}, one finds the massless pole to be
\beqa
A_s&=&\beta\pi\Tr(\epsilon_3\cdot\epsilon_4)\frac{u't'}{s}\,,\labell{As}
\eeqa
while all other terms in \reef{ATG} are contact terms. One might conclude that this produces a four-momentum contact term for two tachyons and two gravitons, which would be in conflict with our field theory action \reef{fourmax}, which has no coupling between two gravitons and two tachyons. We will see shortly that there is in fact no inconsistency between the string amplitude and the field theory amplitude.

It has been shown in \cite{Garousi:2003db} that the leading-order action \reef{Sbulk2} produces the following \(s\)-channel contribution:
\beqa
A_s'&=&-i\kappa^2\Tr(\epsilon_3\cdot\epsilon_4)\Big[\frac{(u-t)^2-s^2}{16 s}+\frac{s}{4}\Big]\,.
\eeqa
This action also produces the following contact term at two-derivative order:
\beqa
A'_c&=&\frac{i\kappa^2}{4}\Tr(\epsilon_3\cdot\epsilon_4)s\,.
\eeqa
Thus, at two-derivative order, field theory yields
\beqa
A_s'+A_c'=\frac{i\kappa^2}{4}\Tr(\epsilon_3\cdot\epsilon_4)\frac{u't'}{s}\,.
\eeqa
With the string amplitude normalization \(\beta = i\kappa^2/(4\pi)\), the above leading amplitude exactly matches the string amplitude \reef{As}.

We now consider the contact terms of the string amplitude at quadratic order in the Mandelstam variables. Using the expansions \reef{exT} and \reef{exG}, one finds
\beqa
A_c^{(2)}&=&-\pi\beta\Tr(\epsilon_3\cdot\epsilon_4)u't'\,.\labell{Ac}
\eeqa
Field theory has no couplings between two tachyons and two gravitons; however, the leading-order action \reef{Sbulk2} and the four-derivative action \reef{fourmax} produce a non-zero \(s\)-channel amplitude for two tachyons and two gravitons. On the other hand, the string amplitude has no massless \(s\)-channel pole other than \reef{As}, which we have already considered. Hence, the field-theory \(s\)-channel amplitude should reduce to contact terms only.

To explicitly compute this amplitude in field theory, we first transform to the Einstein frame using the relation \(G_{\mu\nu} = e^{\gamma\Phi} g_{\mu\nu}\), where \(\gamma = 4/(D-2)\). In this frame, the leading-order action takes the form
 \beqa
 \bS_{\rm leading} &=& \frac{1}{\kappa^2}\int d^{26}x \, \sqrt{-g} \left[ 2\left( R - \gamma \partial_\mu \Phi \partial^\mu \Phi \right) - \frac{1}{2}\partial_\mu T \partial^\mu T -2 e^{\gamma\Phi} V(T) \right] \,. \labell{Sbulkbrane}
 \eeqa
and the four-derivative action takes the form
 \beqa
{\bf S}_{\mathrm{M}}^{(1)}&=&\frac{2}{4\kappa^2}\int d^{26}x \sqrt{-g}\Big[ e^{-\gamma\Phi}( R_{\alpha \beta \gamma 
\delta } R^{\alpha \beta \gamma \delta }  - 4 
R_{\alpha \beta } R^{\alpha \beta }+ R^2)+\cdots\Big]\,.\labell{fourmax1}
\eeqa
where the dots represent other terms that are not relevant for the field-theory amplitude we are interested in. We need standard kinetic terms for the fluctuations, so we normalize the fields as
 \beqa
 G_{\mu\nu} = \eta_{\mu\nu} + \kappa h_{\mu\nu}, \qquad \Phi = \frac{\kappa}{2\sqrt{\gamma}} \phi, \qquad T = \kappa \tau\,.
 \eeqa
The \(s\)-channel contribution in field theory up to order \(\alpha'\), i.e., from \(\bS = \bS_{\rm leading} + \alpha' \bS^{(1)}\), is given by
\beqa
 A'^{(2)}_s &=& \tilde{V}_{\tau_1\tau_2\phi} \tilde{G}_\phi \tilde{V}_{\phi h_3h_4} + (\tilde{V}_{\tau_1\tau_2h})^{\mu\nu} (\tilde{G}_h)_{\mu\nu\alpha\beta} (\tilde{V}_{h h_3h_4})^{\alpha\beta}\,, \labell{A's}
 \eeqa
where the propagators and the first vertex in each term are extracted from the leading-order action \reef{Sbulkbrane}, while the second vertices are extracted from the four-derivative action \reef{fourmax1}. They are given by
\beqa
 \tilde{G}_\phi &=& -\frac{i}{k^2}, \qquad (\tilde{G}_h)_{\alpha\beta,\mu\nu} = -\frac{i}{2k^2} \left( \eta_{\alpha\mu}\eta_{\beta\nu} + \eta_{\alpha\nu}\eta_{\beta\mu} - \frac{2}{D-2}\eta_{\alpha\beta}\eta_{\mu\nu} \right), \nn\\
 \tilde{V}_{\tau_1\tau_2\phi } &=& i\kappa\sqrt{\gamma}, \qquad (\tilde{V}_{ \tau_1\tau_2 h})^{\mu\nu} = -\frac{i\kappa}{2} \left[ p_1^\mu p_2^\nu + p_2^\mu p_1^\nu - \eta^{\mu\nu} (p_1\cdot p_2 + 2) \right]\,,\nn\\
 \tilde{V}_{\phi h_3h_4}&=&-i\kappa\sqrt{\gamma}(p_3\cdot p_4)^2\Tr(\epsilon_3\cdot\epsilon_4)\,,\nn\\
 (\tilde{V}_{h h_3h_4})^{\alpha\beta}&=&-2i\kappa\Big[\Tr(\epsilon_3\cdot\epsilon_4)\left(p_3\cdot p_4 p_3^{(\alpha}p_4^{\beta)}-\frac{1}{2}(p_3\cdot p_4)^2\eta^{\alpha\beta}\right)+\epsilon_3^{(\alpha}\epsilon_4^{\beta)}(p_3\cdot p_4)^2\Big]\,.\nn
 \eeqa
Note that our convention is \(\alpha' = 2\). Substituting these expressions into \reef{A's}, one finds the first and second terms on the right-hand side of \reef{A's} to be
\beqa
 \tilde{V}_{\tau_1\tau_2\phi} \tilde{G}_\phi \tilde{V}_{\phi h_3h_4}&=&-\frac{2i\kappa^2}{D-2}(p_3\cdot p_4)\Tr(\epsilon_3\cdot\epsilon_4)\,,\nn\\
  (\tilde{V}_{\tau_1\tau_2h})^{\mu\nu} (\tilde{G}_h)_{\mu\nu\alpha\beta} (\tilde{V}_{h h_3h_4})^{\alpha\beta}&=&-\frac{i(D-4)\kappa^2}{(D-2)}(p_3\cdot p_4)\Tr(\epsilon_3\cdot\epsilon_4)\nn\\&&+\frac{i\kappa^2}{2}\Tr(\epsilon_3\cdot\epsilon_4)(p_1\cdot p_4 p_2\cdot p_3+p_1\cdot p_3 p_2\cdot p_4-p_3\cdot p_2 p_1\cdot p_2)\,.\nn
 \eeqa
Using the on-shell relations \(p_1\cdot p_3 = p_2\cdot p_4 = -1 - u/2\), \(p_1\cdot p_4 = p_2\cdot p_3 = -1 - t/2\), \(p_1\cdot p_2 = -2 - s/2\), and \(p_3\cdot p_4 = -s/2\), one finds that the amplitude \reef{A's} becomes
\beqa
  A'^{(2)}_s&=&\frac{i\kappa^2}{8}\Tr(\epsilon_3\cdot\epsilon_4)(8-s^2+4t+t^2+4u+u^2)\nn\\
  &=&-\frac{i\kappa^2}{4}\Tr(\epsilon_3\cdot\epsilon_4)t'u'\,,
 \eeqa
where in the last line we have used the on-shell relation \reef{stu2T2G}. This exactly matches the string contact term at quadratic order in the Mandelstam variables \reef{Ac}, using the normalization \(\beta = i\kappa^2/(4\pi)\) already fixed by the massless pole analysis. This agreement confirms our prescription for expanding the tachyon amplitude, as well as the field-theory couplings \reef{fourmax} obtained from T-duality.

 \section{Conclusion}
 
In this paper, we have initiated the study of the higher-derivative effective action of bosonic string theory for tachyon and massless fields using T-duality, combined with a recently proposed expansion of S-matrix elements for massless and tachyon vertex operators \cite{Garousi:2026nvz}. According to this proposal, the effective action should contain only couplings with an even number of tachyons, and the expansion should be such that the leading-order terms are reproduced by the kinetic terms of the action. To impose T-duality, one must first construct a basis of independent couplings and then impose the T-duality constraint to fix the corresponding coupling constants. We have observed that there exists a scheme in which the tachyon without derivatives appears only in the tachyon potential, while all other couplings involve covariant derivatives of the tachyon and the massless fields. Within this scheme, we have constructed a minimal basis at four-derivative order, consisting of 14 independent terms. Imposing T-duality on these couplings reduces them to 6 nonzero couplings, expressed in terms of two parameters. One of these is fixed by comparison with the zero-tachyon case, while the other is determined by comparison with the S-matrix element of four tachyons. The resulting couplings are as follows:

 \beqa
{\bf S}^{(1)}_{\rm MT}&=&\frac{2}{4\kappa^2}\int d^{26}x \sqrt{-G}e^{-2\Phi}\Big[R_{\alpha \beta \gamma \delta } R^{\alpha \beta \gamma \delta }+ \frac{1}{24} H_{\alpha }{}^{\delta \epsilon } H^{\alpha \beta \gamma } H_{\beta \delta }{}^{\varepsilon } H_{\gamma \epsilon \varepsilon } \nn\\&&\qquad\qquad\qquad\qquad\qquad  -  \frac{1}{8} H_{\alpha \beta }{}^{\delta } H^{\alpha \beta \gamma } H_{\gamma }{}^{\epsilon \varepsilon } H_{\delta \epsilon \varepsilon } -  \frac{1}{2} H_{\alpha }{}^{\delta \epsilon } H^{\alpha \beta \gamma } R_{\beta \gamma \delta \epsilon }\nn\\&&\qquad\qquad\qquad\qquad\qquad
-\frac{1}{4}H_\alpha{}^{\gamma\delta}H_{\beta\gamma\delta}\nabla^\alpha T\nabla^\beta T-\frac{1}{16}(\nabla_\alpha T\nabla^\alpha T)^2\Big]\,.\labell{fourmin2}
\eeqa
In the zero-tachyon limit, this reduces to the effective action in the Metsaev–Tseytlin scheme \cite{Metsaev:1987zx}. Using field redefinitions, we have found a form of the action in which the propagators of the leading-order action receive no higher-derivative corrections. In this scheme, the action takes the following form:

 \beqa
{\bf S}_{\mathrm{M}}^{(1)}&=&\frac{2}{4\kappa^2}\int d^{26}x \sqrt{-G}e^{-2\Phi}\Big[ \frac{1}{24} H_{\alpha 
}{}^{\delta \epsilon } H^{\alpha \beta \gamma } H_{\beta 
\delta }{}^{\varepsilon } H_{\gamma \epsilon \varepsilon } -  
\frac{1}{8} H_{\alpha \beta }{}^{\delta } H^{\alpha \beta 
\gamma } H_{\gamma }{}^{\epsilon \varepsilon } H_{\delta 
\epsilon \varepsilon } \nn\\&&+ \frac{1}{144} H_{\alpha \beta \gamma 
} H^{\alpha \beta \gamma } H_{\delta \epsilon \varepsilon } H^{
\delta \epsilon \varepsilon } + H_{\alpha }{}^{\gamma \delta } 
H_{\beta \gamma \delta } R^{\alpha \beta } - 4 
R_{\alpha \beta } R^{\alpha \beta } -  
\frac{1}{6} H_{\alpha \beta \gamma } H^{\alpha \beta \gamma } 
R + R^2 \nn\\&&+ R_{\alpha \beta \gamma 
\delta } R^{\alpha \beta \gamma \delta } -  
\frac{1}{2} H_{\alpha }{}^{\delta \epsilon } H^{\alpha \beta 
\gamma } R_{\beta \gamma \delta \epsilon } -  
\frac{2}{3} H_{\beta \gamma \delta } H^{\beta \gamma \delta } 
\nabla_{\alpha }\nabla^{\alpha }\Phi + \frac{2}{3} H_{\beta 
\gamma \delta } H^{\beta \gamma \delta } \nabla_{\alpha }\Phi 
\nabla^{\alpha }\Phi \nn\\&&+ 8 R \nabla_{\alpha }\Phi 
\nabla^{\alpha }\Phi - 16 R_{\alpha \beta } 
\nabla^{\alpha }\Phi \nabla^{\beta }\Phi + 16 \nabla_{\alpha 
}\Phi \nabla^{\alpha }\Phi \nabla_{\beta }\Phi \nabla^{\beta 
}\Phi - 32 \nabla^{\alpha }\Phi \nabla_{\beta }\nabla_{\alpha 
}\Phi \nabla^{\beta }\Phi \nn\\&&+ 2 H_{\alpha }{}^{\gamma \delta } 
H_{\beta \gamma \delta } \nabla^{\beta }\nabla^{\alpha }\Phi+\frac{1}{8}(\nabla_\alpha T\nabla^\alpha T)^2\Big]\,.\labell{fourmax2}
\eeqa
In the zero-tachyon limit, the action takes the Meissner-scheme form \cite{Meissner:1996sa}. This version of the action is amenable to direct comparison with the expansion of string-theory S-matrix elements. It indicates the absence of any coupling between two tachyons and two gravitons, which we have verified is exactly consistent with the string-theory amplitude.
 
The comparison of the effective action with the expansion of the S-matrix element of four tachyon vertex operators also fixes the \(T^4\) term in the tachyon potential, which appears in \reef{VT}. Interestingly, the coefficient of the \(T^4\) term is positive. This indicates that although bosonic string theory around the trivial Minkowski background with \(T=0\) is unstable, with tachyon mass \(m^2 = -4/\alpha'\), the classical  theory is stable around the minimum of the potential. A closed form for this potential around the unstable point may be
\[
V_{\rm Max}(T) = \frac{2}{\alpha'}(-1 + \cos T).
\]
If this is the case, then the minimum of the potential would occur at \(T = \pi\). At that point, the tachyon becomes a massive scalar with mass \(m^2 = 4/\alpha'\). By shifting the tachyon as \(T \to T + \pi\), one can move from unstable point to the stable point. This shift does not affect the effective actions we have found, except for the tachyon potential, which now becomes
\[
V_{\rm Min}(T) = \frac{2}{\alpha'}(-1 - \cos T).
\]
Around the minimum, the classical effective action admits an Anti–de Sitter solution with cosmological constant \(\Lambda = -4/\alpha'\).
On the other hand, the divergent torus-level dilaton tadpole in bosonic string theory can be canceled by adding a  torus-level cosmological constant to the effective action \cite{Fischler:1986ci,Fischler:1986tb}. This may change \(\Lambda = -4/\alpha'\) to different values—possibly even to positive values\footnote{The one-loop cosmological term in type 0 theory, which also receives contributions from the tachyon, is positive \cite{Baykara:2026gem}.}—which would then transform the Anti–de Sitter solution into a de Sitter solution.
 To provide justification for the above proposal for the tachyon potential, one requires the \(T^6\) term in the tachyon potential \reef{V}. It would therefore be interesting to extend the four-derivative effective action found in this paper to six-derivative order, which might determine, among other things, the \(T^6\) term in the tachyon potential.

It has been observed that when the classical effective action includes a cosmological constant, all higher-derivative couplings can be absorbed into this term via an appropriate higher-derivative field redefinition \cite{Codina:2023fhy}. This does not imply, however, that there is no effective action around the minimum of the tachyon potential with a nonzero cosmological constant. To clarify this, we note that the classical effective action of string theory is background independent \cite{Garousi:2022ovo}, whereas its loop corrections are not \cite{Garousi:2025gsl}. Consequently, the classical effective actions \reef{Sbulk2} and \reef{fourmin2}, which we derived from T-duality and S-matrix elements around the maximum of the tachyon potential, remain valid at the minimum of the potential as well. The background-independent effective action \reef{Sbulk2} contains no classical cosmological constant. Thus, the tachyon potential in this action is \reef{VT}, which has no cosmological constant. Only the effective action at the minimum acquires a cosmological constant; the background-independent potential \reef{VT}, valid for both the maximum and the minimum, does not.

We have imposed that the T-duality corrections satisfy \(\delta \bar{T}=0\) and the relation \reef{dV1} to prevent \reef{dV} from producing non-derivative tachyon terms in the base space. This is required for the scheme in which the non-derivative tachyon appears only in the tachyon potential. The constraint \reef{dV1} requires the coefficient of \(a_8\) in the basis \reef{L1} to be zero. Hence, the tachyon propagator does not receive higher-derivative corrections. One may alternatively constrain  \reef{dV} to produce no non-derivative tachyon couplings by requiring the corrections \(\delta \bar{g}_{ab}\) and \(\delta \bar{\phi}\) to be arbitrary; however, the non-derivative terms produced by the first term in  \reef{dV} would then be canceled by an appropriate correction for \(\delta \bar{T}\). In that case, the coefficient \(a_8\) would remain arbitrary. However, for consistency with the S-matrix element, one must impose the condition that the tachyon propagator does not receive corrections. 
In fact, if one allows the propagator to receive corrections, the field theory in the Meissner scheme receives the coupling \(\nabla_\alpha \nabla_\beta T \nabla^\alpha \nabla^\beta T\) with coefficient \(a_8\), as well as a coupling between two tachyons and two \(B\)-fields with the same coefficient. The presence of the first term produces a massless \(s\)-channel contribution to the amplitude of two tachyons and two gravitons, which is not consistent with the expansion of the string theory amplitude in \reef{As}. Hence, the coefficient of the term \(\nabla_\alpha \nabla_\beta T \nabla^\alpha \nabla^\beta T\) must be zero.

We have seen that, due to the tachyon potential terms, there exists a scheme in which the tachyon without derivatives appears only in the tachyon potential. Type 0 theory also has a tachyon potential, so it seems that the same should hold for the effective action of type 0 theory. In the literature \cite{Klebanov:1998yya,Garousi:2003db}, however, the effective action contains a \(T^2 F^2\) term, where \(F\) is the field strength of the R-R field. This term arises because one assumes the coupling \(T F \bar{F}\) to be present in the effective action. However, it has been argued in \cite{Garousi:2026nvz} that it is consistent to exclude such couplings from the effective action. In present paper, we show that such term can be removed by applying an appropriate field redefinition of the tachyon field to the tachyon mass term.  

To show that without including the coupling \(T F \bar{F}\) there would also be no \(T^2 F^2\) coupling, consider the S-matrix element of two tachyons with \(m^2 = -1\) and two massless R-R vertex operators, which is given in \cite{Klebanov:1998yya,Garousi:2003db}:
 \beqa
 A&\sim& 2\pi\alpha \frac{\Gamma(-u/2)\Gamma(-s/2)\Gamma(-t/2)}{\Gamma(1+u/2)\Gamma(1+s/2)\Gamma(1+t/2)}\,,
 \eeqa
where \(\alpha\) is a kinematic factor that involves the R-R polarization and the four external momenta  \cite{Klebanov:1998yya,Garousi:2003db}. The Mandelstam variables satisfy \(s+t+u=-2\). If one does not include the coupling \(T F \bar{F}\) in the effective action, field theory can produce only the massless pole in the \(s\)-channel. Therefore, the expansion of the S-matrix element should be taken around \(s \to 0\), with \(t,u \to -1\). Defining the new Mandelstam variables as \(t' = t+1=-2p_1\cdot p_4\) and \(u' = u+1=-2p_1\cdot p_2\), the above amplitude becomes
 \beqa
 A&\sim& 2\pi\alpha \frac{\Gamma(1/2-u'/2)\Gamma(-s/2)\Gamma(1/2-t'/2)}{\Gamma(1/2+u'/2)\Gamma(1+s/2)\Gamma(1/2+t'/2)}\,.
 \eeqa
The expansion should be taken around \(s, t', u' \to 0\). The expansion of the Gamma functions is \(-2/s + \log 16 - s \log^2 4 + \cdots\). The leading term of the amplitude is then fully reproduced by the standard kinetic terms of the tachyon and R-R fields  \cite{Klebanov:1998yya,Garousi:2003db}, while all remaining terms produce four- and higher-momentum couplings. Hence, there would be no coupling  \(T^2 F^2\) at two derivative order. 

Another justification for omitting the \(T F \bar{F}\) coupling from the effective action is that the sphere-level S-matrix element of four R-R vertex operators with different chiralities is nonzero (see Appendix in \cite{Garousi:2003db}). To reproduce this amplitude in field theory, one must send all Mandelstam variables to zero, since the external states are massless and satisfy the on-shell relation \(s+t+u=0\). Therefore, the tachyon pole in this amplitude—which would be reproduced by field theory if it included \(T F \bar{F}\)—must be expanded. Since it is expanded, field theory should not contain the \(T F \bar{F}\) coupling. We thus expect that there exists a scheme in which the leading-order effective action of type 0 theory takes the following form:
\beqa
 \bS_{\rm type 0} &=& \frac{1}{\kappa^2}\int d^{10}x \, \sqrt{-g} \, e^{2\Phi} \left[ 2\left( R + 4\partial_\mu \Phi \partial^\mu \Phi - \frac{1}{12}H_{\mu\nu\alpha} H^{\mu\nu\alpha} \right) - \frac{1}{2}\partial_\mu T \partial^\mu T + \frac{1}{2}T^2 \right]\nn\\ &&
  -\frac{1}{\kappa^2}\int d^{10}x \, \sqrt{-g} \left[\frac{1}{2}\sum_{n=1}^4(F_{(n)}\cdot F_{(n)}+\bar{F}_{(n)}\cdot \bar{F}_{(n)})+\frac{1}{2}F_{(5)}\cdot F_{(5)}\right]\,,\labell{type0}
 \eeqa
where the R-R terms with \(n=2,4\) correspond to type 0A, while the remaining R-R terms correspond to type 0B theory. In type 0 theory, there are no \(T^4\) corrections to the tachyon potential \cite{Garousi:2003db}. 

It has been conjectured that M-theory on the wedge space \(S^1 \vee S^1\) \cite{Baykara:2026gem} corresponds to the strong-coupling limit of type 0A string theory\footnote{The Hořava–Witten theory formulated on this space has also been considered in \cite{Altavista:2026evd,Altavista:2026brr}.}. This requires the type 0 tachyon to admit a geometric interpretation, relating it to the radii of the two circles in the wedge space \(S^1 \vee S^1\) \cite{Baykara:2026gem,Dasgupta:2026maq,Basile:2026trt,Kamal:2026msr}. Tachyon condensation to \(T = 2\), which corresponds to shrinking one of the circles to zero radius, transforms type 0A theory into type IIA \cite{Baykara:2026gem}. This condensation eliminates the \(\bar{F}_2\) and \(\bar{F}_4\) fields, and the vacuum energy associated with the tachyon mass term is expected to be canceled by the loop-level cosmological constant \cite{Baykara:2026gem}. Under this tachyon condensation, the above classical effective action for the type 0A case reduces to the bosonic part of the type IIA effective action.

Although the corrections to the R-R couplings in \reef{type0} begin at four-derivative order, the higher-derivative corrections to the tachyon and massless NS-NS fields in the first line above start at eight-derivative order \cite{Garousi:2003db}. It would be interesting to use T-duality to determine the higher-derivative corrections to the above type 0 effective action.

\end{document}